\documentclass[12pt, draftclsnofoot, onecolumn]{IEEEtran}
\usepackage[utf8]{inputenc}

\usepackage{amsmath}
\usepackage{amssymb}
\usepackage{amsfonts}
\usepackage{cite}

\usepackage[caption=false]{subfig}
\usepackage{booktabs}
\usepackage{bbm}
\usepackage{mathrsfs}

\DeclareMathOperator*{\argmax}{arg\,max}

\newcommand{\Rmnum}[1]{\expandafter\@slowromancap\romannumeral #1@}
\usepackage{graphicx}

\usepackage{epsfig}
\usepackage[numbers,sort&compress]{natbib}
\usepackage{xcolor}
\usepackage{color}
\usepackage{enumerate}
\usepackage{extarrows}
\usepackage[ruled,linesnumbered,lined]{algorithm2e}  
\usepackage{algorithmic}
\usepackage{tensor}
\usepackage[colorlinks,linkcolor=blue,anchorcolor=blue,citecolor=blue,urlcolor=black]{hyperref}
\usepackage{url}

\begin{document}

\title{Neuromorphic Non-Orthogonal Multiple Access for Parallel Remote Inference via  Vector Symbolic Architecture}
\author{Jiechen Chen,  \IEEEmembership{Member,~IEEE}, Zihang Song, \IEEEmembership{Member,~IEEE},  Dengyu Wu, \IEEEmembership{Member,~IEEE}, Bipin Rajendran,~\IEEEmembership{Senior Member,~IEEE}, and \\Osvaldo Simeone,~\IEEEmembership{Fellow,~IEEE}
\thanks{J. Chen and D. Wu are with the Department of Engineering, King’s College London, London, WC2R 2LS, UK (email:\{jiechen.chen, dengyu.wu\}@kcl.ac.uk). Z. Song is with the Department of Electronic Systems, Aalborg University, 9220 Aalborg, Denmark (e-mail:zsong@es.aau.dk). B. Rajendran and O. Simeone are with the Institute for Intelligent Networked Systems,
Northeastern University London, One Portsoken Street, London, E1 8PH,
UK (email: \{b.rajendran, o.simeone\}@nulondon.ac.uk).\\
This work was supported by the European Research Council (ERC) under the European Union’s Horizon Europe Programme (grant agreement No. 101198347), by an Open Fellowship of the EPSRC (EP/W024101/1), and by the EPSRC project (EP/X011852/1).
 }
\vspace*{-1.5cm}
}

\maketitle

\begin{abstract}
Emerging edge intelligence systems increasingly rely on dense deployments of always-on sensors that must convey task-relevant information to a remote model under tight energy and spectral budgets.  The deployment of event-driven neuromorphic sensing paired with spiking neural networks (SNNs) is attractive in this regime because it produces dynamically sparse representations, so that energy is spent on communication and computation only when informative events occur. Prior multiple-access protocols for remote inference using neuromorphic sensing and computing  targeted {collaborative} settings, in which the server fuses information from all devices into a {single} decision. This paper instead addresses {parallel} remote inference, in which each device observes a {distinct} input, and requires its own classification decision. We propose NOMA-NC, a non-orthogonal multiple-access (NOMA) neuromorphic communication (NC)  protocol built on the vector symbolic architecture (VSA) framework. In NOMA-NC, each device binds its sparse spike feature map with a device-specific permutation key, and all devices in a group transmit concurrently so that the over-the-air superposition directly realizes the VSA bundling operation. A shared decoding SNN, together with lightweight per-device learned unbinding, recovers all decisions in a single inference pass. Experiments on the N-MNIST and DVS128 Gesture datasets show that NOMA-NC yields goodput gains and savings in terms of receiver computing energy  that are sub-proportional to the number of simultaneously active devices, without increasing the per-device transmission energy.
\end{abstract}

\begin{IEEEkeywords}
Neuromorphic computing, spiking neural networks, non-orthogonal multiple access, vector symbolic architecture.
\end{IEEEkeywords}

\IEEEpeerreviewmaketitle
\newtheorem{definition}{\underline{Definition}}[section]
\newtheorem{fact}{Fact}
\newtheorem{assumption}{Assumption}
\newtheorem{theorem}{\underline{Theorem}}[section]
\newtheorem{lemma}{\underline{Lemma}}[section]
\newtheorem{proposition}{\underline{Proposition}}[section]
\newtheorem{corollary}[proposition]{\underline{Corollary}}
\newtheorem{example}{\underline{Example}}[section]
\newtheorem{remark}{\underline{Remark}}[section]
\newcommand{\mv}[1]{\mbox{\boldmath{$ #1 $}}}
\newcommand{\mb}[1]{\mathbb{#1}}
\newcommand{\Myfrac}[2]{\ensuremath{#1\mathord{\left/\right.\kern-\nulldelimiterspace}#2}}
\newcommand\Perms[2]{\tensor[^{#2}]P{_{#1}}}
\newcommand{\note}[1]{[\textcolor{red}{\textit{#1}}]}

\section{Introduction} \label{intro}

\subsection{Motivation}
Next-generation wireless edge intelligence systems are expected to support dense deployments of low-cost sensors that continuously feed a shared, capable model hosted at an edge server. In such systems, the classical objective of reliably reconstructing transmitted bits is wasteful: what matters is not the fidelity of the raw measurements, but the quality of the downstream decisions they enable. This observation has motivated the shift toward goal-oriented and task-oriented communications, in which the transmitter, the channel access strategy, and the receiver are co-designed around the task to be accomplished rather than around bit-level reconstruction \cite{popovski2020semantic,strinati2021goal, gunduz2023beyond}. When many sensors share the spectrum and the remote model, the dominant costs become the wireless resources consumed in the uplink and the computational energy expended at the edge server, both of which must be controlled jointly with task performance.

Event-driven neuromorphic sensing is a natural enabler for energy-efficient always-on monitoring systems. Sensors such as \emph{event-driven cameras} respond only to changes in the scene, emitting asynchronous, binary events rather than dense frames \cite{4444573, lenero20113}. Furthermore,  \emph{spiking neural networks} (SNNs) \cite{8259423, davies2021advancing, 10606014} -- or more broadly neuromorphic processing units \cite{simeone2026modern, lee2025spiking} -- can efficiently extract information from  event data. In particular, benefiting from the  dynamic sparsity of the sensed data, neuromorphic processors can ensure that, at any time step, only a small fraction of neurons fire.  Overall, because energy is consumed only when an event or a spike occurs, an event-driven sensing-and-communication pipeline communicates and computes \emph{only when needed} for the task, translating the intrinsic sparsity of the data into direct savings in transmission and computation energy \cite{chen2023neuromorphic}. 

\begin{figure}[htp]
    \centering
\includegraphics[width=1.0\linewidth]{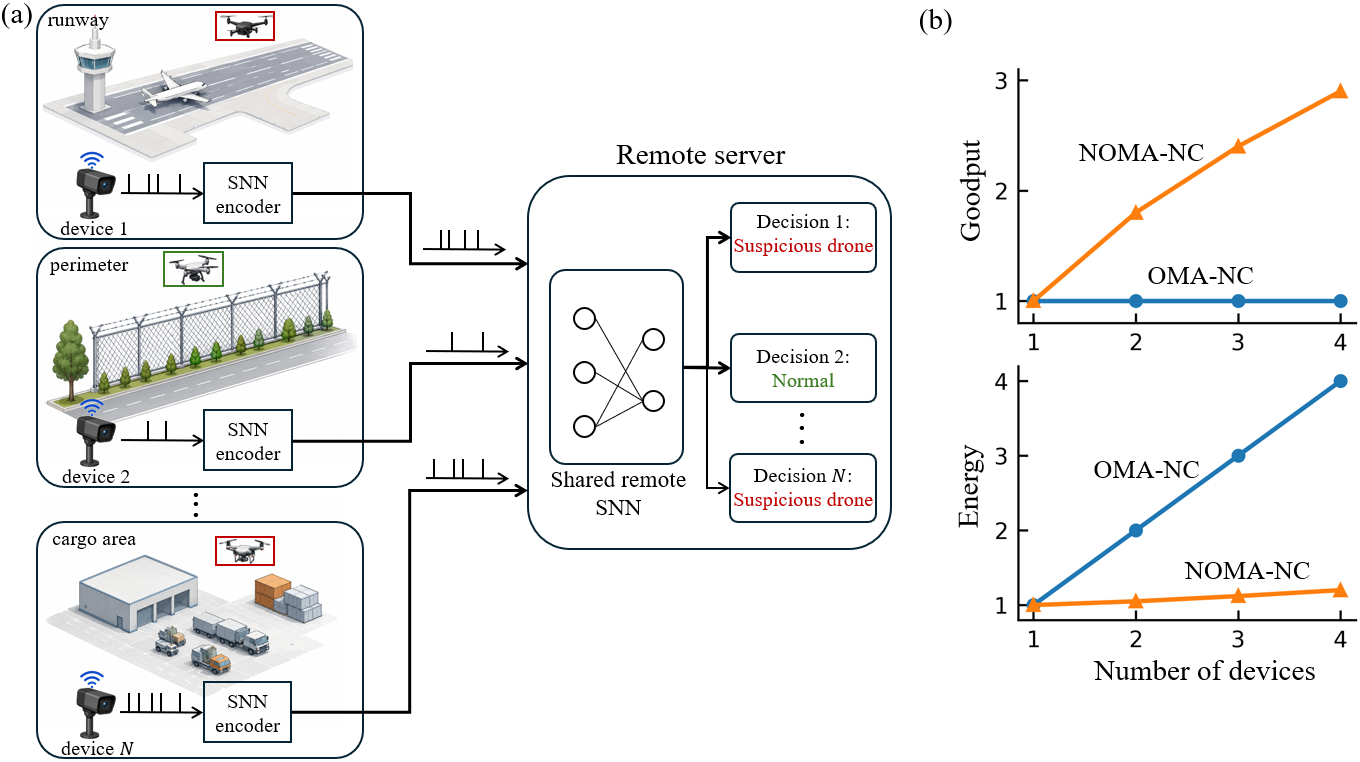}
    \vspace{-1cm}
    \caption{Parallel remote inference for multi-device monitoring via neuromorphic communications (NC). (a) Multiple devices equipped with event-driven sensors monitor distinct regions of interest, such as different sectors of an airport, and encode their local event streams into sparse spike-domain representations for wireless transmission to an edge server. The edge server uses a shared remote SNN to produce independent decisions for the respective devices, e.g., to detect suspicious drone activity in each sector. (b) Qualitative scaling of normalized goodput (average number of correct decisions per frame) and normalized edge-server computing energy versus the number of devices for conventional orthogonal multiple access (OMA) NC, i.e., OMA-NC and the proposed non-orthogonal multiple access (NOMA) NC, i.e., NOMA-NC. Compared with OMA-NC, NOMA-NC supports concurrent transmissions and shared remote inference, improving goodput while reducing the growth of receiver-side computing energy.   }
    \label{gmodel}
\end{figure}

This principle underpins a growing line of work on neuromorphic wireless communications and split computing, in which neuromorphic sensing and on-device SNN encoding are coupled with sparse, event-driven transmission to a remote SNN for remote inference \cite{chen2023neuromorphic, 10682971, 10946192, wu2025neuromorphic, song2025neuromorphic}. Fig. \ref{gmodel}(a) illustrates one such system, in which a number of devices monitor different areas, e.g., different sectors of an airport, and communicate to an edge server  to enable inference on the respective local observations, e.g., to detect suspicious drone activity.

Existing neuromorphic remote inference protocols, however, are  designed for \emph{collaborative} inference: multiple devices observe the same scene or different views of a common phenomenon, and the server fuses their transmissions into a single decision \cite{chen2023neuromorphic, 10682971, 10946192, wu2025neuromorphic}.  As in the example in Fig. \ref{gmodel}(a), many practical deployments instead call for \emph{parallel} remote inference, where each device monitors a distinct region of interest and demands its own, independent decision from the same shared model.  

In this setting, two objectives are simultaneously important. First, the system should sustain a high \emph{goodput}, defined as the number of correct decisions produced per communication frame and spectrum across all devices. Second, the \emph{computational energy} expended by the edge server should remain low, despite the need to serve multiple devices. These goals are challenging, since a conventional implementation based on orthogonal multiple access (OMA) would require spectral and computational resources at the edge server that scale proportionally to the number of devices. In fact, in a conventional system with OMA, the server would need to carry out a separate inference pass for the signal received from each device (see Fig. \ref{gmodel}(b)). 

In this work, we propose a non-orthogonal multiple-access (NOMA) protocol that can obtain a more favorable scaling of goodput and receiver energy consumption {(see Fig. 1(b)).} The approach leverages  the idea of computation in superposition, which was recently put forth by MIMONets \cite{menet2023mimonets}. In MIMONets, multiple inputs are bound into a single composite representation and processed by one shared model, reducing the per-input computational cost. MIMONets  builds on the \emph{vector symbolic architecture (VSA)}, also known as \emph{hyperdimensional computing}, a framework that formalizes binding and bundling operations to combine and decouple high-dimensional vector representations \cite{kleyko2022survey}.

MIMONets, however, considers a \emph{centralized and noiseless} setting in which the multiple inputs are available directly at the processor. An extension to a \emph{remote, noisy} scenario was recently studied in \cite{abdi2025semantic}, which assumes a single source connected to a remote server over a wireless channel. This prior art does not address the setting of interest, which encompasses spatially distributed, energy-constrained devices that must communicate over a noisy, fading wireless channel. Bridging this gap is the central goal of this work.

\subsection{State of the Art}
\subsubsection{Goal-oriented and semantic communications}
The reconceptualization of communication around tasks and semantics rather than bit reconstruction has been articulated in several foundational works \cite{gunduz2023beyond, strinati2021goal}. These frameworks advocate co-designing source encoding, channel access, and inference for the end task, and provide the conceptual backdrop for the present paper. They are, however, largely agnostic to the energy profile of the underlying hardware and do not exploit event-driven sparsity.

\subsubsection{Neuromorphic wireless communications and split computing}
A complementary line of work realizes goal-oriented communication on neuromorphic hardware. Event-driven semantic communication for remote inference was introduced in \cite{chen2023neuromorphic}, and subsequently extended with wake-up radios and digital-twin-aided design \cite{10682971}, multi-level spikes \cite{10946192}, and resonate-and-fire neurons \cite{wu2025neuromorphic}. These protocols demonstrate the energy benefits of transmitting and processing spikes only when they occur. Another line of work \cite{gupta2026neuromorphicrx} places the SNN at the receiver, replacing channel estimation, equalization, and demapping in a 5G-NR OFDM system for energy-efficient bit detection.  Crucially, however, they consider single-input single-output or multi-input single-output configurations, in which the devices either serve a single shared task or are allocated orthogonal time--frequency resources, so that the spectral footprint and the server's inference load grow with the number of devices.

\subsubsection{Over-the-air computation and collaborative inference}
A separate body of work exploits the waveform-superposition property of the wireless multiple-access channel to compute functions of distributed data directly over the air \cite{zhu2019broadband}. This principle has been applied to collaborative edge inference, where the superposed transmissions of many clients are aggregated into the features or beliefs feeding a single inference decision \cite{yilmaz2025private, wen2022task}. Such schemes are spectrally efficient and can preserve model privacy, but they fundamentally produce \emph{one} decision from the contributions of all devices, and they do not consider event-driven sensing or the device- and receiver-side energy constraints that arise with neuromorphic hardware.

\subsubsection{Vector symbolic architectures and computation in superposition}
VSA represents and manipulates information using high-dimensional vectors through algebraic binding, bundling, and unbinding operations \cite{plate1995holographic, kanerva2009hyperdimensional, kleyko2022survey, kleyko2022vector}. Because randomly chosen high-dimensional vectors are quasi-orthogonal with high probability, multiple bound items can be superposed in a single fixed-width vector and later retrieved with low interference \cite{frady2018theory}. For example, in holographic
reduced representations (HRR), binding is implemented as circular convolution between two $D$-dimension vectors $\mv x$ and $\mv a$, while in the multiply-add-permute (MAP) model binding is the
element-wise (Hadamard) product of bipolar vectors.

MIMONets \cite{menet2023mimonets} leverages this property for computation in superposition, passing several bound inputs through one neural network to reduce per-input compute. As noted above, this is a centralized, noiseless construction. Semantic multiplexing \cite{abdi2025semantic} extends this idea to a wireless scenario, binding multiple task representations at a single transmitter and sending the resulting composite signal to an edge server over a MIMO channel. In contrast, our work operates over a multiple-access channel, where spatially distributed neuromorphic devices transmit concurrently and the bundling is realized physically by the over-the-air superposition of their spiking signals.

\subsection{Contributions}
This paper proposes a neuromorphic NOMA protocol, termed NOMA-NC,  for parallel remote inference. As shown in Fig. \ref{model}, in NOMA-NC  multiple devices with distinct inputs are served by a single shared model through over-the-air VSA bundling. The main contributions are summarized as follows.

\begin{itemize}
    \item \emph{Problem formulation:} We formulate parallel remote inference over a shared wireless multiple-access channel with neuromorphic, event-driven devices, and characterize the system through three coupled metrics: the goodput, defined as the number of correct decisions per receiver inference; the per-device transmission energy; and the receiver-side compute energy, measured in synaptic operations on neuromorphic hardware \cite{8259423}.

    \item \emph{NOMA-NC:} We propose a VSA-based non-orthogonal multiple-access protocol in which each device binds its sparse spike feature map with a device-specific permutation key. The binding is folded directly into the OFDM subcarrier mapping as a form of index modulation, incurring no additional device-side computation and preserving spike sparsity, so that the per-device transmission energy matches that of the orthogonal baseline. Concurrent transmission realizes the VSA bundling operation over the air, and a shared decoding SNN with lightweight per-device learned unbinding recovers the decisions for all devices in a group in a \emph{single} inference pass, breaking the linear growth of receiver compute energy with the number of devices (see Fig. \ref{gmodel}(b)).

    \item \emph{Hyper-NOMA-NC:} To remove the assumption of transmitter-side channel inversion, we introduce a receiver-side mechanism in which a lightweight hypernetwork maps the received pilot observation to a per-channel scaling vector that compensates the fading prior to decoding. Equalization and decision making are optimized jointly end-to-end, so that channel feedback to the devices is not required and the devices remain maximally lightweight.

    \item \emph{Experimental validation:} On the N-MNIST dataset and DVS128 Gesture dataset,  the proposed schemes scale the goodput approximately linearly with the group size $G$, substantially reduce the receiver-side compute energy relative to the orthogonal baseline, and does not increase the per-device transmission energy, confirming that the additional capability of non-orthogonal parallel inference does not burden the devices.
\end{itemize}

\subsection{Organization}
The remainder of this paper is organized as follows. Section~II reviews the necessary background on vector symbolic architectures, including the binding, bundling, and unbinding operations adopted in this work. Section~III presents the system model, covering the neuromorphic sensing and encoding pipeline, the transmission model, and the design goals expressed through goodput, transmission energy, and receiver compute energy. Section~IV develops the proposed NOMA-NC protocol under transmitter-side channel inversion, detailing permutation-based binding, over-the-air bundling, receiver-side unbinding and inference, and the training procedure. Section~V introduces Hyper-NOMA-NC, which performs joint receiver-side equalization and decision making via a hypernetwork without channel feedback. Section~VI reports the experimental results, and Section~VII concludes the paper.

\section{Background: Vector Symbolic Architecture}
Since the proposed neuromorphic NOMA protocol builds on VSA, this section presents a brief review of VSA following reference \cite{kleyko2022survey}. VSA represents information using high-dimensional vectors, which are composed, combined, and decoupled through algebraic operations known as binding, bundling, and unbinding \cite{kleyko2022survey, kleyko2022vector}. 
\subsubsection{Binding}
Binding ties together two information vectors into an individual coherent representation. Mathematically, it maps two $D$-dimensional vectors to a third vector of the same dimension, producing a representation that is quasi-orthogonal to both vectors, thus constituting an independent information vector. For example, consider encoding the concept of a blue ball.  We assign independent random vectors $\mv v_{\rm ball} \in \mathbbm{R}^D$ and $\mv v_{\rm blue} \in \mathbbm{R}^D$ to the object and its color, respectively. Denoting the binding operation as $\odot$, binding them yields
\begin{align}
    \mv v = \mv v_{\rm ball} \odot \mv v_{\rm blue},
\end{align}
where vector $\mv v$ is quasi-orthogonal to both vectors $\mv v_{\rm ball}$ and $\mv v_{\rm blue}$, forming an independent representation of the blue ball. Apart from being dimensionality-preserving, the binding operation $\odot$ is approximately invertible, enabling the estimate of the original vectors from the bound vector. Furthermore, it distributes over addition, supporting combination, or bundling, as discussed below. 

Several concrete realizations of binding have been proposed in VSA \cite{kleyko2022survey}, and in this work we adopt \emph{permutation-based binding}. This binds a real-valued vector $\mv x \in \mathbb{R}^D$ to an integer vector $\mv a \in [D]^D$ selected as one of the $D!$ permutations of the integers $[D]=\{1, \ldots, D\}$. Specifically, denoting as $\Pi_{D}$ the set of permutations of set $[D]$, vector $\mv x \in \mathbb{R}^D$ is bound to a permutation $\mv a=[a_1, \ldots, a_D]\in \Pi_D$ by rearranging its elements according to the fixed permutation $\mv a$, i.e.,
\begin{align}
    \mv x \odot \mv a = \mv P_{ \scalebox{0.8}{\mv a}} \, \mv x, \label{bind}
\end{align}
with $\mv P_{ \scalebox{0.8}{\mv a}}$ the $D \times D$ permutation matrix whose $(i, j)$-th entry is defined as 
\begin{equation}
\mv P_{ \scalebox{0.8}{\mv a}}^{ij}=\left\{
\begin{array}{ll}
1,        & \text{if}~ j=a_i, \\
0,       &   \text{otherwise}.
\end{array} \right. 
\end{equation}

When the permutation $\mv a$ is selected uniformly at random from set $\Pi_D$, the bound vector $\mv P_{ \scalebox{0.8}{\mv a}}\mv x$ becomes quasi-orthogonal to the original vector $\boldsymbol{x}$ with high probability. To see this, write the inner product between $\mv x$ and $\mv P_{ \scalebox{0.8}{\mv a}}\mv x$ as $\mv x^\top \mv P_{ \scalebox{0.8}{\mv a}}\mv x = \sum_{i=1}^D x_i x_{a_i}$. Since each $a_i$ is marginally uniformly distributed over the $D$ positions, taking the expectation over the random permutation yields $\mathbb{E}[\mv x^\top \mv P_{ \scalebox{0.8}{\mv a}}\mv x]=(\sum_{i=1}^D x_i)^2/D$. For a vector whose entries sum to approximately zero, this mean is negligible, and a similar calculation shows that the variance is of order $\mathcal{O}(1/D)$ by the concentration of measure phenomenon \cite{kanerva2009hyperdimensional,frady2018theory}. Hence  $\mv x^\top \mv P_{ \scalebox{0.8}{\mv a}}\mv x$ concentrates around zero as the dimension $D$ grows, so $\mv P_{ \scalebox{0.8}{\mv a}}\mv x$ is nearly orthogonal to $\mv x$.

\subsubsection{Bundling}
Given $N$ $D$-dimensional vectors $\mv z_1, \ldots, \mv z_N$, bundling combines them into a single composite representation of the same dimension. The standard bundling operation is element-wise addition. Accordingly, when the individual vectors $\mv z_n$ are bound as in \eqref{bind}, i.e., as $\mv z_n=\mv x_n \odot \mv a_n$ for some permutation $\mv a_n \in \Pi_D$, bundling yields the superposition
\begin{align}
    \mv z =\sum_{n=1}^{N} \mv z_{n}. \label{super}
\end{align}

\subsubsection{Unbinding} \label{unbinding}
When the constituent vectors $\mv z_n$ are quasi-orthogonal, the bundled representation $\mv z$ in \eqref{super} retains retrievable information about each component, with inter-component interference diminishing as the dimensionality $D$ grows. Recovery is done using the unbinding operator. In practice, given the bundle $\mv z$ in \eqref{super}, unbinding with the permutation $\mv a_n$ retrieves an approximation of the corresponding vector $\mv x_n$ together with a noise term arising from the remaining components. Denoting the unbinding operator as $\circledast$, the approximate recovery of vector $\mv x_n$ yields
\begin{align}
    \hat{\mv x}_n=\mv a_n \circledast \mv z = \mv x_n + \underbrace{\mv a_n \circledast \sum_{n^{\prime} \neq n}\mv x_{n^{\prime}} \odot \mv a_{n^{\prime}}}_{\text{noise}}. \label{noise}
\end{align}
For permutation binding, the unbinding operator is often chosen as the inverse permutation matrix $\mv P_{ \scalebox{0.8}{\mv a}}^{-1}=\mv P_{\bar{ \scalebox{0.8}{\mv a}}}$ \cite{kleyko2022survey}, where $\bar{\mv a}$ is the inverse permutation of $\mv a$. Substituting the permutation unbinding operator into \eqref{noise} gives
\begin{align}
    \hat{\mv x}_n = \mv P_{\scalebox{0.8}{$\bar{\mv{a}}_n$}} \mv z = \mv x_n + \sum_{n' \neq n} \mv P_{\scalebox{0.8}{$\bar{\mv{a}}_n$}}\mv  P_{\scalebox{0.8}{$\mv{a}_{n^{\prime}}$}} \mv x_{n'},
\end{align}
where each noise vector $\mv P_{\scalebox{0.8}{$\bar{\mv{a}}_n$}} \mv P_{\scalebox{0.8}{$\mv{a}_{n^{\prime}}$}} \mv x_{n'}$ consists of a random rearrangement of the entries of vector $\mv x_{n'}$. Since the keys $\mv a_n$ and $\mv a_{n^{\prime}}$ are drawn independently and uniformly at random, the composite permutation $\mv P_{\scalebox{0.8}{$\bar{\mv{a}}_n$}}\mv  P_{\scalebox{0.8}{$\mv{a}_{n^{\prime}}$}}$ is itself a uniformly random permutation, so each noise term is quasi-orthogonal to $\mv x_n$, with high probability for large dimension $D \gg N$.

Given that for finite dimension $D$, the residual noise in \eqref{noise} may be non-negligible, one can replace the inverse binding operations $\mv P_{\scalebox{0.8}{$\bar{\mv{a}}_n$}}$ with a learned operation within some class.   For example, unbinding can be implemented via a linear projection $\mv A_n \in \mathbb{R}^{D \times D}$ trained to extract the $n$-th component from the bundle by minimizing the squared error $\| \mv x_n - \mv A_n \mv z \|^2 $ \cite{kleyko2022survey}.

\section{System Model}
\begin{figure}[htp]
    \centering
\includegraphics[width=1.0\linewidth]{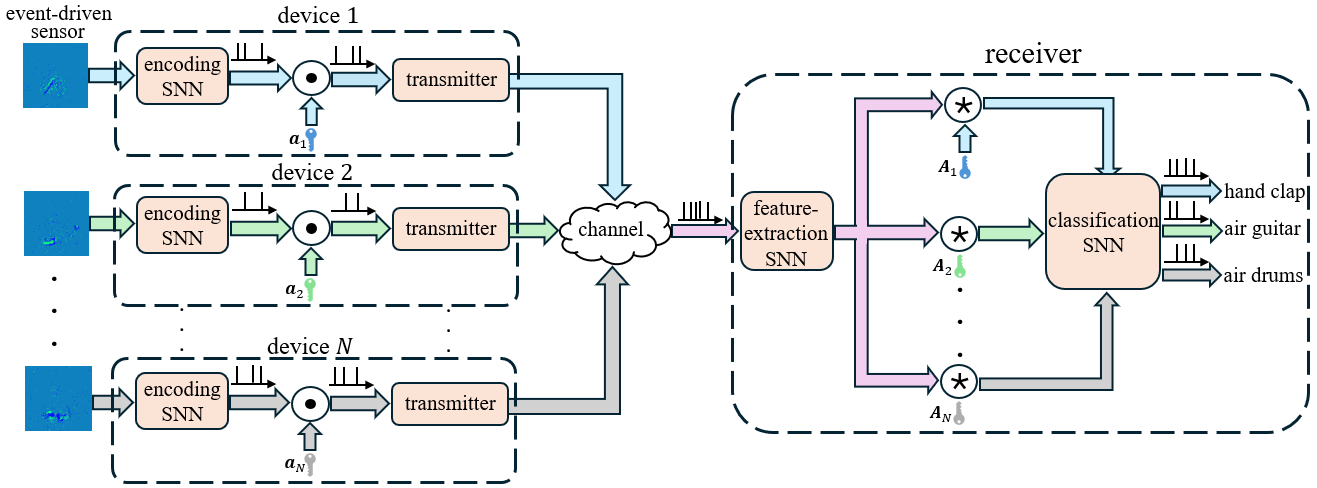}
    \vspace{-1cm}
    \caption{In the proposed NOMA-NC protocol, each of the $N$ devices processes its local event stream through an  encoding SNN, and binds the resulting spike feature map with a user-specific permutation key $\mv a_n$. The bound representations are transmitted simultaneously over a shared wireless channel, where the over-the-air superposition naturally realizes the bundling operation in VSA. At the edge server, the decoding SNN processes the received signal using the learned unbinding matrices $\mv A_n$, and then a shared spiking output layer produces the classification decision simultaneously for all devices.   }
    \label{model}
\end{figure}

In this section, we present the system illustrated in Fig. \ref{model}, where multiple devices, equipped with neuromorphic sensors,  share a wireless channel for transmission to an edge server for remote inference. Unlike the neuromorphic split computing systems studied in \cite{chen2023neuromorphic, 10682971, 10946192, wu2025neuromorphic}, in which all devices observe the same input and share the same inference task, the devices observe distinct inputs requiring separate inference decisions from a shared model. As discussed in Sec. \ref{intro}, a possible use case is airport surveillance, in which each event-driven camera monitors a distinct sector, e.g., the runway, the perimeter, and the cargo area, and the shared model at the edge server independently flags suspicious drone activity in each sector.

The main design goal is improving over the baseline implementation in which each device operates on orthogonal frequency resources by leveraging the neuromorphic protocol and using the shared model separately for each device. Specifically, we aim at enhancing goodput, via the use of a non-orthogonal (NOMA) protocol based on VSA, and computing efficiency, by sharing a single forward pass of the shared model to produce decision for all devices simultaneously.

We start by describing the setting, then the neuromorphic sensing and encoding model, followed by the transmission model, and finally formalize the design goal.

\subsection{Neuromorphic Sensing and Encoding}

As shown in Fig. \ref{model}, we consider a system in which $N$ devices collect data from an event-driven sensor \cite{4444573}, such as  iniVation DAVIS event camera \cite{6889103}, and  communicate with an edge server over a shared wireless channel. All $N$ devices aim to perform the same $C$-class classification task on the local inputs by leveraging the model at the receiver. 

\subsubsection{Sensing Model}
Unlike conventional frame-based cameras that record intensity values, an event-driven camera responds to changes in brightness at each pixel, producing a stream of binary events. Specifically, an output equal to 0 (no spike) indicates a static pixel; an output equal to +1 (positive spike) indicates a positive change; and an output equal to -1 (negative spike) indicates a negative change. Following standard practice \cite{10242251}, the event stream is discretized into $T$ time steps, yielding a sequence of ``spiking" images $\mv U_{n,1}, \ldots, \mv U_{n,T}$, i.e., tensors $\mv U_{n,t} \in\{0,1\}^{2 \times H \times W}$ with binary entries. Each tensor $\mv U_{n,t}$ represents the output of the event-driven camera at device $n$ and time step $t$, with $H$ and $W$ denoting the camera's resolution in terms of number of pixels along the height and width dimension, respectively, and the two channels corresponding to positive and negative polarity events.  

\subsubsection{Encoding SNN}
Each device encodes its spiking image sequence through a local encoding SNN over the $T$ time steps. The use of an SNN ensures that computing benefits from dynamic sparsity, consuming energy only on parts of the input containing a spike \cite{8259423}. In a manner compatible with most neuromorphic computing platforms \cite{schuman2017survey}, the encoding SNN is composed of leaky integrate-and-fire (LIF) neurons.  Accordingly, each neuron maintains a scalar state variable $u_t$ along the time index $t=1,2, \ldots$, and emits a spike when its state $u_{t}$ exceeds a firing threshold $\vartheta$, i.e.,
\begin{align}
    s_{t} =\Theta(u_{t} -\vartheta), \label{heavi}
\end{align}
with $\Theta(\cdot)$ denoting the Heaviside step function. 
Given a set $\mathcal{N}$ of neurons feeding into the current neuron, the state $u_t$ evolves according to the dynamics
\begin{align}
    u_{t} = \beta u_{t-1} + \sum_{j\in\mathcal{N}}w_{j} x_{j,t} -s_{t-1} \vartheta, \label{membrane}
\end{align}
where $\beta \in(0,1)$ is the membrane leak factor controlling the decay of past information; $w_{j}$ is the synaptic weight from neuron $j \in \mathcal{N}$; and $x_{j,t}$ is the input from neuron $j\in\mathcal{N}$. Because the inputs $x_{j, t}$ are binary spikes, the weighted sum $\sum_{j\in\mathcal{N}}w_{j} x_{j,t}$ reduces to accumulating the weights $w_j$ of only those presynaptic neurons that fire at time $t$, and is therefore evaluated using ADD operations alone without multiplications. The term $s_{t-1} \vartheta$  implements the reset mechanism, subtracting the threshold $\vartheta$ from the state whenever the neuron fired at the previous time step.

Following the LIF model \eqref{heavi}-\eqref{membrane},  the output of the encoding SNN at each time step $t$ is a latent spike feature map  $\mv S_{n,t} \in\{0,1\}^{D \times H^{\prime} \times W^{\prime}}$, where $H^{\prime}$ and $W^{\prime}$ are the spatial dimensions of the feature map after encoding, and $D$ is the dimension of the feature vector $s_{n,t}^{h,w} \in \{0,1\}^D$ learned by the encoding SNN at each spatial location $(h,w)$ with $h\in\{1, \ldots, H^{\prime}\}~\text{and}~ w\in\{1, \ldots, W^{\prime}\}$. The dimensions $H^{\prime}$ and $W^{\prime}$ are smaller than the input dimensions, i.e., $H^{\prime} <H$ and $W^{\prime} <W$, as a result of the spatial downsampling performed by the encoding SNN, while the dimension $D$ is larger than the two input polarity channels, i.e., $D>2$, reflecting the expansion into a richer set of feature channels at each spatial location. Ideally, the spike feature maps $\mv S_{n,t}$ produced by the encoding SNN is sparse, including a small fraction of non-zero entries.

\subsection{Transmission Model} \label{transmission}
The devices transmit their spike feature maps $\mv S_{n,t}$, $n=1, \ldots, N$, to the edge server over a wireless channel. In prior work \cite{chen2023neuromorphic, wu2025neuromorphic}, two options were considered for transmission of binary, i.e., spiking signals like the feature map $\mv S_{n,t}$, namely ultra-wideband (UWB) \cite{yang2004ultra} and OFDM with index modulation \cite{bacsar2015multiple}. Both types of modulation can benefit from the sparsity of the input $\mv S_{n,t}$ by allocating transmission energy  only to the spikes, i.e., to the non-zero elements of tensor $\mv S_{n,t}$. In this work, in order to facilitate multiple access protocols from devices to server, we adopt OFDM with index modulation.

\begin{figure}[htp]
    \centering
\includegraphics[width=0.9\linewidth]{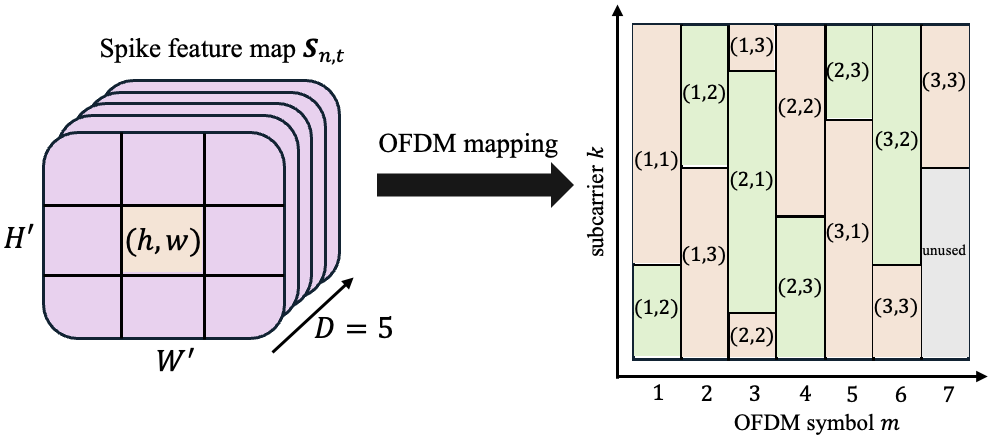}
    \caption{An illustration of OFDM resource grid layout of the spike feature map. Each spatial location $(h,w)$ of the spike feature map $\mv S_{n,t} \in \{0,1\}^{D \times H^{\prime} \times W^{\prime}}$ is assigned a contiguous block of $D$  resource elements, with blocks tiled sequentially across the resource grid and possibly spanning OFDM symbol boundaries. The example illustrates the case $H^{\prime} = W^{\prime}=3$ and $D=5$ over a resource grid with $K=7$ subcarriers, requiring $M=7$  OFDM symbols (see \eqref{number}).    }
    \label{map}
\end{figure}

As illustrated in Fig. \ref{map}, if each OFDM symbol has $K$ data-bearing subcarriers, the number of OFDM symbols required to transmit a single spike feature map $\mv S_{n,t}$ is
\begin{align}
    M = \left\lceil \frac{H'W'D}{K} \right\rceil, \label{number}
\end{align}
where $\lceil \cdot \rceil$ denotes the ceiling function, which rounds up to the nearest integer. Following index modulation \cite{bacsar2013orthogonal}, within the $M$ OFDM symbols, only the subcarriers corresponding to non-zero latent features are activated, while the remaining subcarriers are idle. As a result, the set of active subcarriers within each block directly encodes the firing pattern of the corresponding spike vector. Owing to the dynamic sparsity of the spike feature maps, only a small fraction of the $KM$ resource elements carry non-zero values.

We write the transmitted signals from the $n$-th device across all $K$ subcarriers of the $m$-th OFDM symbol as the vector $\mv x_{n,t}^m =[x_{n,t}^{m,1}, \ldots, x_{n,t}^{m,K}] \in \mathbb{C}^K$.  The transmitted signal $x_{n,t}^{m,k} \in \mathbb{C}$ of device $n$ on the $k$-th subcarrier of the $m$-th OFDM symbol is given by
\begin{align}
    x_{n,t}^{m,k} \in \{0, \sqrt{P}\}, \label{power}
\end{align}
where $P$ denotes the maximum per-subcarrier transmission power. Accordingly, if $\rho_{n,t} \in [0,1]$ is the fraction of active spiking elements in the transmission signal $\mv X_{n,t}=[\mv x_{n,t}^1, \ldots, \mv x_{n,t}^M]$  of device $n$ at time $t$, the overall transmission power of device $n$ at time $t$ is $\rho_{n,t} P$. This modulation scheme thus translates the spiking sparsity of the latent representation directly into a reduction in transmit energy consumption.

\subsubsection{Channel Model}
The wireless channel between each device $n$ and the edge server is modeled as a frequency-selective multi-path fading channel with $L$ taps that remains constant across the $M$ OFDM symbols allocated to one time step $t$.  The cyclic prefix of each OFDM symbol is assumed longer than the channel delay spread, so that the time-domain convolution reduces to a per-subcarrier multiplication in the frequency domain. Furthermore, it is also long enough to accommodate any residual frame synchronization at the level of OFDM symbols.

The channel frequency response for device $n$ on the $k$-th subcarrier is given by
\begin{align}
    h_{n}^{k} = \sum_{i=1}^L \alpha_{n}^i \exp\!\left(-j \frac{2\pi}{K} \tau_{n}^i k\right),
\end{align}
where $\alpha_{n}^i$ and $\tau_{n}^i$ denote the complex gain and delay of the $i$-th path, respectively \cite{wunder2006delay}. We collect the channel frequency responses across all $K$ subcarriers into the diagonal matrix $\mv H_{n} \in \mathbb{C}^{K \times K}$, whose $k$-th  diagonal entry is the complex channel gain $h_{n}^{k}$. 

\subsubsection{Orthogonal Transmission} \label{omat}
In a conventional baseline implementation, the $N$ devices access the channel through orthogonal multiple access (OMA) via time-division multiple access (TDMA).   With this protocol, at each time $t$, the $m$-th received OFDM signal from device $n$ is  
\begin{align}
    \mv y_{n,t}^m = \mv H_{n} \mv x_{n,t}^m + \mv z_{n,t}^m, \label{oma}
\end{align}
where $\mv z_{n,t}^m \in \mathbb{C}^K$ represents additive noise with i.i.d. complex Gaussian elements having zero mean and variance $N_0$. Since each device must be allocated $M$ OFDM symbols, OMA requires $NM$ OFDM symbols to serve all $N$ devices.
\subsubsection{Non-Orthogonal Transmission} \label{nomat}
The proposed scheme relies on non-orthogonal (NOMA) transmission. With NOMA, a subset $\mathcal{G} \subseteq \{1, \ldots, N\}$ of the $G=|\mathcal{G}|$ devices transmit concurrently over the same  $M$ OFDM symbols. Accordingly, the received signal at the edge server for the $m$-th OFDM symbol at time $t$ is given by the superposition
\begin{align}
    \mv y_t^m = \sum_{n \in \mathcal{G}} \mv H_{n} \mv x_{n,t}^m + \mv z_t^m \label{channel} 
\end{align}
of the transmitted signals from all devices. By allowing for the simultaneous transmission from $G$ devices, NOMA requires $\lceil N/G \rceil \cdot M$ OFDM symbols, reducing to the minimum allocation of $M$ OFDM symbols when all devices transmit concurrently, i.e., $G=N$. 

\subsubsection{Pilot Transmission}
Each block of $M$ OFDM symbols corresponding to a time step $t$ is preceded by a sounding phase in which every device transmits a known pilot vector $\mv p_n \in \mathbb{C}^K$ over the $K$ subcarriers of one dedicated OFDM symbol. In OMA, the pilots are transmitted on disjoint resources, yielding the per-device received pilot
\begin{align}
    \mv y_n^{\rm pilot} = \mv H_n \mv p_n + \mv z_n^{\rm pilot};
\end{align}
while in NOMA, the $G$ devices within one group transmit their pilots simultaneously, yielding the superposed received pilot
\begin{align}
    \mv y^{\rm pilot} = \sum_{n \in \mathcal{G}} \mv H_n \mv p_n + \mv z^{\rm pilot}. \label{nomap}
\end{align}

Overall, the number of pilot OFDM symbols for OMA is equal to the number of devices, $N$, while for NOMA with group size $G$, we need $\lceil N/G \rceil$ pilot OFDM symbols.

\subsection{SNN Decoder}
At the edge server, an SNN decoder processes the received signal and produces $N$ classification decisions, one for each device. Under the OMA baseline, the decoder processes each device on its own dedicated resources and therefore runs $N$ separate inference passes, one per device, to produce the $N$ decisions. Under the proposed NOMA scheme, detailed in Sec. IV, the decoder instead operates on the superimposed signal from a group of devices and recovers all of their decisions within a single shared inference pass, which is the key to the receiver-side computational savings targeted in this work.

\subsection{Design Goal} \label{goal}
We characterize the performance of the system through three metrics, namely goodput, transmission energy, and computing energy. Let $P_{\rm acc}\in[0,1]$ denote the probability of successful classification, i.e., the average fraction of correct classification decisions made at the server across all devices.

\emph{Goodput:} The goodput is defined as the average number of correct classification decisions produced per block of $M$ OFDM symbols, i.e.,
\begin{align}
    \mathcal{T} = \frac{G P_{\rm acc}}{1+1/M}, \label{through}
\end{align}
where we recall that $G$ is the number of devices transmitting simultaneously over $M$ OFDM symbols. The numerator is the average number of correct decisions per group, and the denominator $(M+1) /M$ accounts for the pilot overhead, which vanishes as $M$ grows. Increasing the group size $G$ thus can potentially enhance the goodput, as long as the accuracy $P_{\rm acc}$ is not excessively degraded by the simultaneous transmissions of $G$ devices.

\emph{Transmission energy:}  The transmission energy  $E_{\rm tx}$ is defined as the average transmission energy expended per device and per subcarrier as 
\begin{align}
    E_{\rm tx} = \frac{\sum_{t=1}^T \sum_{n=1}^N \rho_{n,t} P}{TN}. \label{transmitp}
\end{align}

\emph{Computing energy:} In SNNs implemented on digital neuromorphic hardware, all synaptic updates reduce to addition operations \cite{8259423}. Accordingly, we count the number of synaptic ADD operations of the SNN in order to evaluate the computing energy. Since all schemes adopt the same SNN architecture for the devices, we specifically focus on the energy consumption of the decoding SNN at the server in order to compare the protocols under study.

To elaborate, let $L_t^{\rm ADD}$ denote the number of synaptic ADD operations performed by the decoding SNN at time $t$. The receiver-side computing energy is then evaluated as the average energy per device per time step, i.e.,
\begin{align}
    E_{\rm rx} = \frac{\epsilon\sum_{t=1}^T L_t^{\rm ADD}}{TG}, \label{rxe}
\end{align}
where $\epsilon$ is the energy per ADD operation, e.g., $\epsilon=23.6$ pJ for Intel Loihi \cite{8259423}. The metric \eqref{rxe} accounts for the fact that, with  groups of size $G$, one pass on the $T$ time steps of the received data is sufficient to make classification decisions for $G$ devices. Therefore, the per-time step energy $\epsilon \sum_{t=1}^T L_t^{\rm ADD}/T$ is shared across $G$ devices. Increasing the group size $G$ thus reduces the receiver's energy consumption metric \eqref{rxe}, but at the potential cost of a degraded goodput. 

The goal of this work is to design a NOMA protocol that increases the goodput $\mathcal{T}$, while reducing the receiver-side computing energy $E_{\rm rx}$, for a fixed transmission energy $E_{\rm tx}$.

\section{NOMA-Neuromorphic Communications}
In this section, we present the proposed NOMA-NC protocol, which leverages the VSA framework, reviewed in Sec. II, to enable non-orthogonal multiple access for the parallel remote inference setting discussed in Sec. \ref{goal}. As illustrated in Fig.~\ref{model}, the $G$ devices in a group superpose their transmissions over the same time-frequency resources, each binding its spike feature map with a unique permutation key, so that the over-the-air superposition realizes VSA bundling and the edge server applies learned unbinding before the shared decoding SNN.

In order to focus on the essential elements of NOMA-NC, this section assumes perfect channel knowledge at the devices. This allows each device to pre-equalize its transmitted signal, creating an effective additive Gaussian noise multi-access channel  \cite{elsawy2014stochastic}.  A receiver-side mechanism that does not require channel knowledge at the devices, and abides by the per-spike power constraint $P$ in \eqref{power}, is then presented in Section V.

\subsection{Permutation-Based Binding at Devices}
The proposed NOMA-NC follows the NOMA transmission strategy presented in Sec. \ref{nomat}, in which the $G$ devices in the same group $\mathcal{G}$ transmit simultaneously over the same $M$ OFDM symbols. Within each group, following VSA (see Sec. II), devices distinguish their transmissions through device-specific permutation keys. Specifically, each device $n$  is assigned a unique permutation key $\mv a_n = [a_{n,1}, \ldots, a_{n,D}] \in \Pi_D$, where we recall that $\Pi_D$ is the set of all permutation of the integers $\{1, \ldots,D\}$. Prior to transmission, device $n$ binds its local spike feature map $\mv S_{n,t}$ with its permutation key $\mv a_n$  independently at each spatial location $(h, w)$ of the spike feature map, with $h\in\{1, \ldots, H^{\prime}\}~\text{and}~ w\in\{1, \ldots, W^{\prime}\}$.
 
 Specifically, denoting by $\mv s_{n,t}^{h,w} \in \{0,1\}^D$ the $D$-dimensional spike vector corresponding to location $(h,w)$, the bound spike vector is given by the permuted sequence
\begin{align}
    \mv {u}_{n,t}^{h,w} = \mv P_{\boldsymbol{a}_n} \mv s_{n,t}^{h,w}, 
\end{align}
where $\mv P_{\boldsymbol{a}_n}$ is the $D \times D$  permutation matrix corresponding to permutation $\mv a_n$. The same permutation is applied at every spatial location and every time step, producing the bound spike feature map $\mv U_{n,t}=[\mv u_{n,t}^{1,1}, \ldots, \mv u_{n,t}^{H^{\prime},W^{\prime}}] \in \{0,1\}^{D \times H^{\prime} \times W^{\prime}}$. 

This position-wise design rearranges only the feature dimension at each spatial location without mixing information across pixels, preserving the spatial locality that the server-side decoding SNN relies on. Permutation binding is particularly well-suited to binary spike representations, since permutations rearrange entries without altering their values, preserving both the spike domain and the sparsity structure of the original spike maps. We note that, while the approach is related to interleave-division multiple access (IDMA) \cite{ping2006interleave}, IDMA aims at multi-user detection, whereas the proposed scheme targets remote inference.

\subsection{Permutation-Based Subcarrier Mapping and Over-the-Air Bundling}
A key practical advantage of permutation-based binding is that it can be folded directly into the OFDM subcarrier mapping, requiring no arithmetic at the device beyond index remapping during modulation. Accordingly, the transmit power expended by each device is given by \eqref{transmitp}, where the sparsity $\rho_{n,t}$ of the transmission signal $\mv X_{n,t}$ equals the fraction of non-zero elements of the feature map  $\mv S_{n,t}$.

As anticipated, in this section we study an idealistic scenario in which the fading channel is pre-compensated at the devices. Under perfect channel knowledge, device $n$ applies channel-inversion pre-compensation, scaling its spike symbol of magnitude $\sqrt{P}$ from \eqref{power} by $1/h_n^k$ on subcarrier $k$, so that the transmitted signal is $\sqrt{P}/h_n^k$ per active spike. The channel $\mv H_n$ in \eqref{channel} is thereby inverted, and the received signal reduces to
\begin{align}
    \mv y_t^m = \sum_{n \in \mathcal{G}} \mv x_{n,t}^m + \mv z_t^m, \label{receive}
\end{align}
thus following a Gaussian multi-access channel.

In this setting, the edge server estimates the ideal bundle $\mv B_t = \sum_{n=1}^G \mv U_{n,t}$ from the received signal \eqref{receive} by reversing the subcarrier mapping, i.e.,
\begin{align}
    \hat{\mv B}_t= \sum_{n \in \mathcal{G}} \mv U_{n,t}+\tilde{\mv Z}_t, \label{nosiyb}
\end{align}
where $\tilde{\mv Z}_t$ is the noise remapped to the corresponding feature and spatial positions. 
The over-the-air superposition thus directly realizes the VSA bundling operation.

\subsection{Receiver-Side Unbinding and Inference}
As shown in Fig. \ref{model}, the estimated $\hat{\mv B}_t$ in \eqref{nosiyb} is first passed through a shared feature-extraction SNN, producing an output spike vector $\mv g_t \in \{0,1\}^{D^{\prime}}$  that contains mixed information from all $G$ devices. The device-specific representations are then recovered using a learned unbinding matrix $\mv A_n \in \mathbb{R}^{D^{\prime} \times D^{\prime}}$ for each device $n$, yielding
\begin{align}
    \hat{\mv g}_{n,t} = \mv A_n \mv g_t. \label{unbinding}
\end{align}
The learned unbinding matrices $\{\mv A_n\}_{n=1}^G$ are ideally designed to jointly compensate for the channel noise, the nonlinear distortion of the decoding SNN, as well as for the residual interference from the remaining $G-1$ devices in the group (see \eqref{noise}). Moreover, since the decoding SNN output $\mv g_t$ is a binary spike vector, the matrix-vector product $\mv A_n \mv g_t$ in \eqref{unbinding} reduces to accumulating the columns of $\mv A_n$ indexed by the non-zero entries of $\mv g_t$. The unbinding is therefore carried out using ADD operations alone, so that it inherits the event-driven energy efficiency of the spiking pipeline.

Each extracted vector $\hat{\mv g}_{n,t}$  is then passed through a shared classification SNN, which emits a spike vector $\mv o_{n,t} \in\{0,1\}^C$ at each time step $t$. After all $T$ time steps have been processed, the classification decision for each device $n$ is obtained by rate decoding \cite{10242251}, accumulating spike counts over time $t$ into the spike count vector $\mv r_n =[r_{n,1}, \ldots, r_{n, C}]$, i.e., \begin{align}
    \mv r_n = \sum_{t=1}^T \mv o_{n,t}. \label{scount}
\end{align}
Following standard practice in rate-coded SNN training \cite{10242251}, we convert the spike counts into a probability distribution $\mv p_n=[p_{n,1}, \ldots, p_{n,C}]$ over the $C$ classes via a softmax function, with the $c$-th element given by
\begin{align}
    p_{n,c} = \frac{\exp(r_{n,c})}{\sum_{c'=1}^C \exp(r_{n, c^{\prime}})}. \label{probability}
\end{align}
The predicted class is the neuron with the highest spike probability, i.e.,
\begin{align}
    \hat{y}_n = \argmax_{c \in \{1,\ldots,C\}} p_{n,c}. \label{decision}
\end{align}

A key advantage of this receiver architecture is that all $G$ devices in the group are decoded through a single forward pass of the shared feature-extraction SNN and the shared classification SNN, with only the lightweight per-device multiplication by the unbinding matrices $\{\mv A_n\}_{n=1}^G$ being applied $G$ times.

\subsection{Training}
The encoding SNNs, feature-extraction SNN, unbinding matrices, and final classification SNN are jointly trained end-to-end under noiseless channel conditions for a chosen group size $G$. The same trained model is then deployed in settings with a possibly different group size $G$. To this end, we assume access to a training dataset $\mathcal{D}=\{(\mv X^i, \mv y^i)\}_{i=1}^{|\mathcal{D}|}$, where $\mv X^i=(\mv X_1^i, \ldots, \mv X_G^i)$ consists of $G$ spiking input streams with $\mv X^i_n=(\mv X_{n,1}^i, \ldots, \mv X_{n,T}^i)$, and $\mv y^i=(y_1^i, \ldots, y_G^i)$ contains the corresponding classification labels with $\mv y^i_n \in \{1, \ldots, C\}$. 

The training objective minimizes the sum of cross-entropy losses over all $G$ devices and all samples in the dataset, i.e.,
\begin{align}
    \mathcal{L} = -\frac{1}{|\mathcal{D}|G} \sum_{i=1}^{|\mathcal{D}|} \sum_{n \in \mathcal{G}} \sum_{c=1}^C q_{n,c}^i \log p_{n,c}^i, \label{lossf}
\end{align}
where $\mv p_n^i=[p_{n,1}^i, \ldots, p_{n,C}^i]$ is the probability vector for sample $i$ with each element obtained from \eqref{probability}, and $\mv q_n^i=[q_{n,1}^i, \ldots, q_{n,C}^i] \in \{0,1\}^C$ is the one-hot encoding of the ground-truth label $y_n^i$.  The loss is minimized using stochastic gradient descent (SGD). Since the Heaviside function $\Theta(\cdot)$ in the LIF neuron is non-differentiable, gradients are computed using a surrogate gradient during the backward pass \cite{neftci2019surrogate}. 

For the OMA baselines, we augment the loss in \eqref{lossf} with a Hoyer sparsity regularizer on the transmitted spikes, which encourages the encoding SNN to fire fewer spikes and thus transmit at lower energy without sacrificing accuracy \cite{wu2025neuromorphic}. For NOMA-NC, no such regularizer is added, since the superposition of transmissions implicitly drives the encoding SNN to produce sparser spikes to limit inter-device interference, as shown in the experiments in Sec. \ref{exp}.

After training, the encoding SNNs and permutation keys are deployed at the device side, while the decoding SNN, unbinding matrices, and classification layer are deployed at the edge server \cite{chen2023neuromorphic, 10682971, 10946192, wu2025neuromorphic}.

\section{Hyper-NOMA-NC: Joint Equalization and Decision Making}
The NOMA-NC protocol of Section IV relies on transmitter-side channel-inversion power control, which requires a channel estimate fed back from the edge server and is sensitive to both estimation error and feedback delay. We now propose Hyper-NOMA-NC, which lifts this assumption by jointly performing channel equalization and decision making at the edge server without channel feedback to the devices.

\subsection{Hypernetwork-Based Equalization}
Under NOMA without transmitter-side equalization, each entry of the received bundle $\hat{\mv B}_t$  is a fading-weighted superposition of the corresponding entries from the $G$ devices in the group, as given in \eqref{channel}. The channel coefficients $\{\mv H_{n}\}_{n \in \mathcal{G}}$ therefore distort the magnitudes of the bundle entries in a manner that depends on the current channel realization.
To compensate for this distortion at the receiver, we introduce a per-channel scaling vector $\mv \gamma=[\gamma_1, \ldots, \gamma_D] \in  \mathbb{R}^D$, generated on the fly from the pilot observation, that modulates the bundle before it is processed by the decoding SNN. 

The scaling vector is produced by a lightweight hypernetwork $\Phi$ with parameters $\mv \phi$, which takes input the received pilot, and produces the scaling factor, i.e.,
\begin{align}
    \boldsymbol{\gamma} = \Phi_{\scalebox{0.7}{\mv \phi}}(\mv y^{\rm pilot}).
\end{align}
The hypernetwork can be implemented as a small feedforward neural network. 
The scaled bundle $\tilde{\mv B}_t \in \mathbb{R}^{D \times H^{\prime} \times W^{\prime}}$ at each time step is obtained by rescaling each of the $D$ feature channels of $\hat{\mv B}_t$ by its corresponding gain, i.e.,
\begin{align}
    \tilde{B}_{t, d}^{h,w} = \gamma_d \, \hat{B}_{t, d}^{h,w}.
\end{align}
The scaled bundle $\tilde{\mv B}_t$ is then processed by the decoding SNN, the learned unbinding matrices $\{\mv A_n\}_{n\in\mathcal{G}}$, the shared spiking output layer, and rate decoding, as described in Section IV.

This design preserves the lightweight nature of the devices, since the device-side processing is identical to that of NOMA-NC and no channel feedback is required. By placing the compensation between the channel and the decoding SNN, the hypernetwork further adapts the decoding pipeline jointly with the channel rather than producing a clean equalized signal as an intermediate target, so that equalization and decision making are jointly optimized end-to-end. This allows the system to compensate for the channel together with the residual interference and the nonlinear distortions of the decoding SNN.

\subsection{End-to-End Training}
The training procedure extends that of Section IV.D by including the wireless channel and by adding the hypernetwork parameters $\boldsymbol{\phi}$ and the pilots $\{\mv p_n\}_{n \in \mathcal{G}}$ to the set of trainable parameters. Given a training sample $\mv X^i$ and a channel realization $\mv H=\{\mv H_n\}_{n \in \mathcal{G}}$, passing the sample through the end-to-end system yields the probability vector $\mv p_n^i(\mv H)$ for each device $n \in \mathcal{G}$. The training objective is the cross-entropy loss, additionally averaged over the channel distribution, i.e.,
\begin{align}
    \mathcal{L} = -\mathbb{E}_{\scalebox{0.7}{\mv H}}\!\left[\frac{1}{|\mathcal{D}| G} \sum_{i=1}^{|\mathcal{D}|} \sum_{n \in \mathcal{G}} \sum_{c=1}^C q_{n,c}^i \log p_{n,c}^i(\mv H)\right].
\end{align}
In practice, the expectation over $\mv H$ is approximated via Monte Carlo by sampling a fresh channel realization per training iteration. The resulting loss is minimized using stochastic gradient descent with surrogate gradients, as in Section IV.D. After training, the encoding SNN, the permutation keys, and the learned pilots are deployed at the device side, while the decoding SNN, the unbinding matrices, the classification layer, and the hypernetwork are deployed at the edge server.

\section{Experiments} \label{exp}
In this section, we present experimental results to elaborate on the advantages of the proposed NOMA-NC protocol. We first describe the experimental setting, and then present results on energy consumption and goodput.

\subsection{Setting}
We consider $N=4$ devices and evaluate the proposed scheme  on two neuromorphic benchmarks, namely N-MNIST \cite{orchard2015converting} and DVS128 Gesture \cite{amir2017low} datasets. N-MNIST consists of neuromorphic MNIST handwritten digit recordings with $C=10$  classes captured by a DVS event camera, where each sample is a sequence of binary events of size $2 \times 34 \times 34$, and the dataset contains 60,000 training samples and 10,000 test samples. DVS128 Gesture is a more challenging benchmark recorded with a DVS128 event camera, comprising 1,342 recordings of $C=11$  hand and arm gestures performed by 29 subjects under three illumination conditions, with recordings from the first 23 subjects used for training and those from the remaining 6 for testing. For both datasets, the two channels correspond to positive and negative polarity events. The N-MNIST event stream is discretized into $T=10$  time steps, while for DVS128 Gesture the events are accumulated into time windows to yield up to 35 time steps per recording, and each input frame is spatially downsampled to size $2 \times 48\times 48$.

For N-MNIST, each device is equipped with an encoding SNN consisting of a $5 \times 5$ convolutional layer with $D=12$ output channels, followed by $2 \times 2$ max-pooling and a LIF layer, producing a spike feature map of spatial dimension $H^{\prime} \times W^{\prime} = 15 \times 15$. At the edge server, the decoding SNN consists of a $5 \times 5$ convolutional layer with $32$ output channels, followed by $2 \times 2$ max-pooling, a LIF layer, and a global average pooling layer that reduces the spatial dimension to $1 \times 1$, yielding a $32$-dimensional vector $\mv g_t$ that is processed by per-device unbinding matrices $\mv A_n \in \mathbb{R}^{32 \times 32}$. 

For DVS128 Gesture, the encoding SNN uses a $5 \times 5$ convolutional layer with $D=32$ output channels followed by $2 \times 2$ max-pooling and a LIF layer, producing a spike feature map of spatial dimension $H^{\prime} \times W^{\prime} = 24 \times 24$. The decoding SNN comprises two further convolutional blocks, with 64 and 128 output channels respectively, each followed by $2 \times 2$ max-pooling and a LIF layer, and a global average pooling layer reducing the spatial dimension to $1 \times 1$,  yielding a 128-dimensional vector $\mv g_t$ processed by per-device unbinding matrices $\mv A_n \in \mathbb{R}^{128 \times 128}$. 

In both cases, the output of each $\mv A_n$ is passed through a LIF layer to produce the spiking representation $\hat{\mv g}_{n,t}$, and the shared spiking output layer is a fully-connected mapping to $C$ neurons followed by a LIF layer. The hypernetwork is a feedforward network with one hidden layer of 128 neurons, taking the received pilot signal as input and outputting the per-channel scaling vector $\mv \gamma$. The pilot sequences $\{\mv p_n\}_{n \in \mathcal{G}}$ are initialized with i.i.d. $\mathcal{CN}(0,1)$ entries and jointly optimized with the rest of the architecture during training.

We adopt OFDM transmission with $K=2048$ subcarriers per OFDM symbol. The wireless channel is modelled as Rayleigh block fading, with the frequency-domain channel gains $h_n^k$ drawn as i.i.d. circularly-symmetric complex Gaussian variables. We define the per-spike signal-to-noise ratio (SNR)  as the ratio between the per-subcarrier transmit power and the noise power $N_0$, i.e.,
\begin{align}
    {\rm SNR} =  \frac{P}{N_0}.
\end{align}

\subsection{Benchmarks}
We compare NOMA-NC against the following baselines:
\begin{itemize}
\item \emph{Joint end-to-end training}: The encoding and decoding SNNs are trained end-to-end through the wireless channel, without any explicit channel compensation. This benchmark can be instantiated in either OMA or NOMA mode. This approach  was used, e.g., in \cite{abdi2025semantic} for a semantic multiplexing scheme based on conventional digital signaling.
\item \emph{OMA-NC}: Each device transmits its spike feature map on disjoint time-frequency resources, with the channel handled either through transmitter-side power inversion or through the same hypernetwork-based receiver-side compensation as in the proposed scheme. This approach essentially coincides with the protocol in \cite{chen2023neuromorphic} applied separately to each device.
\item \emph{Shared-inference OMA-NC}: Transmission and channel sounding are orthogonal as in OMA-NC, but the server bundles the per-device compensated feature maps and decodes them in a single shared inference pass. This isolates the receiver-side compute savings of shared inference from the spectral savings of over-the-air bundling, which it forgoes. Accordingly, this protocol effectively applies MIMONets \cite{menet2023mimonets} to the bundled decoded signal. 

\end{itemize}

\subsection{Perfect Channel Compensation}
We first investigate the energy consumption and goodput of NOMA-NC under the ideal setting in which the channel is perfectly compensated via transmitter-side power inversion based on a noiseless channel estimate. As a result, the received signal at the edge server reduces to a clean superposition of the bound spike feature maps with additive Gaussian noise. 

\begin{figure}
    \centering
    \includegraphics[width=1.0\linewidth]{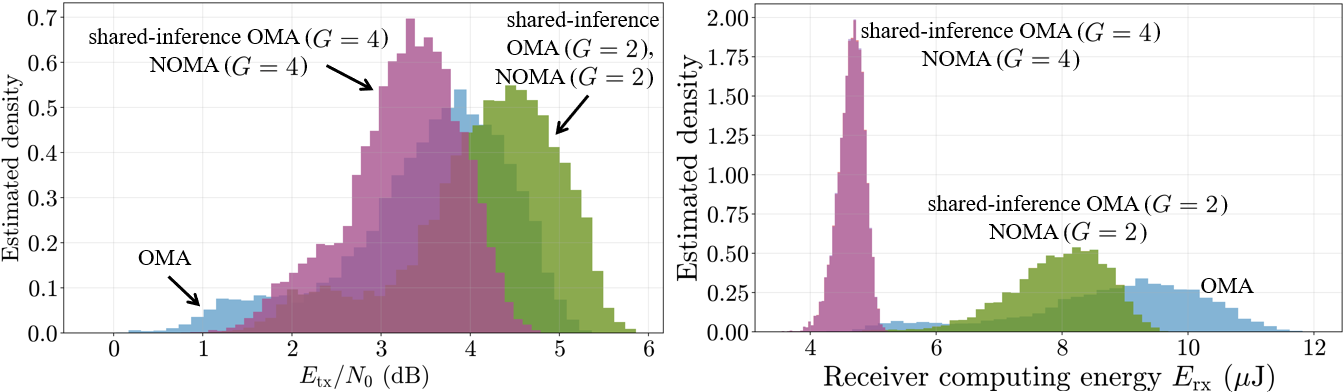}
    \vspace{-1cm}
    \caption{Left: Empirical distribution of the ratio $E_{\rm tx}/N_0$ between transmission energy and noise power. Right: Empirical distribution of the receiver computing energy $E_{\rm rx}$ in \eqref{rxe}. Results are shown for OMA-NC and NOMA-NC with $G \in \{2, 4\}$ and $\text{SNR}=10$ dB for the N-MNIST dataset.
} \label{hist}
\end{figure}

\begin{figure}
    \centering
    \includegraphics[width=0.6\linewidth]{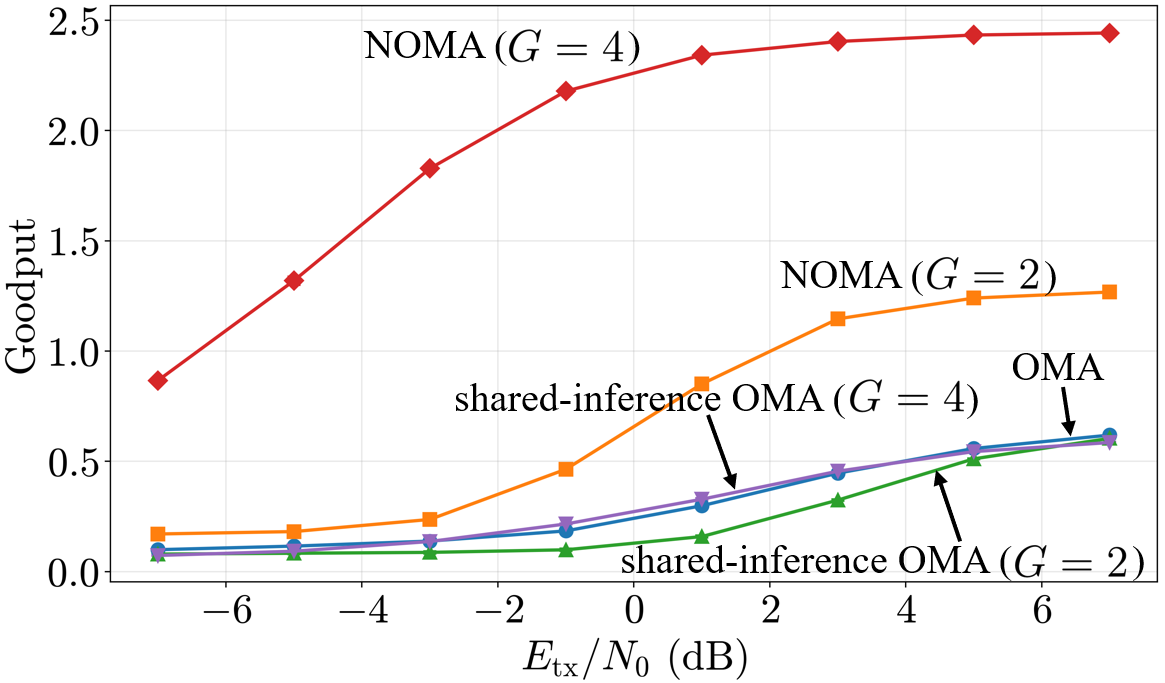}
    \caption{Goodput $\mathcal{T}$ versus $E_{\rm tx}/N_0$ for OMA-NC and NOMA-NC with group sizes $G=2$ and $G=4$ for the N-MNIST dataset.} \label{awgn_snr}
\end{figure}

\begin{figure}
    \centering
    \includegraphics[width=1.0\linewidth]{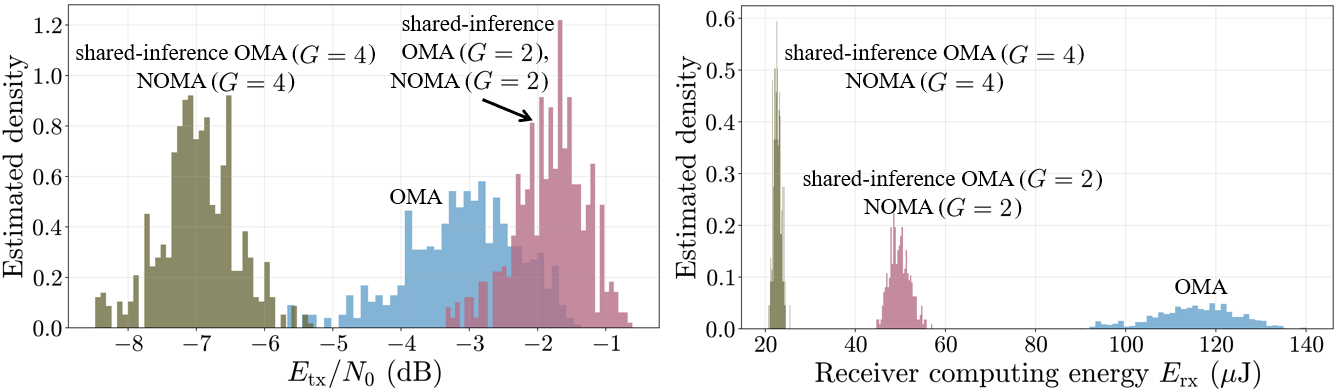}
    \vspace{-1cm}
    \caption{Left: Empirical distribution of the ratio $E_{\rm tx}/N_0$ between transmission energy and noise power. Right: Empirical distribution of the receiver computing energy $E_{\rm rx}$ in \eqref{rxe}. Results are shown for OMA-NC and NOMA-NC with $G \in \{2, 4\}$ and $\text{SNR}=10$ dB for the DVS128 Gesture dataset.
} \label{histnmnist}
\end{figure}

\begin{figure}
    \centering
    \includegraphics[width=0.6\linewidth]{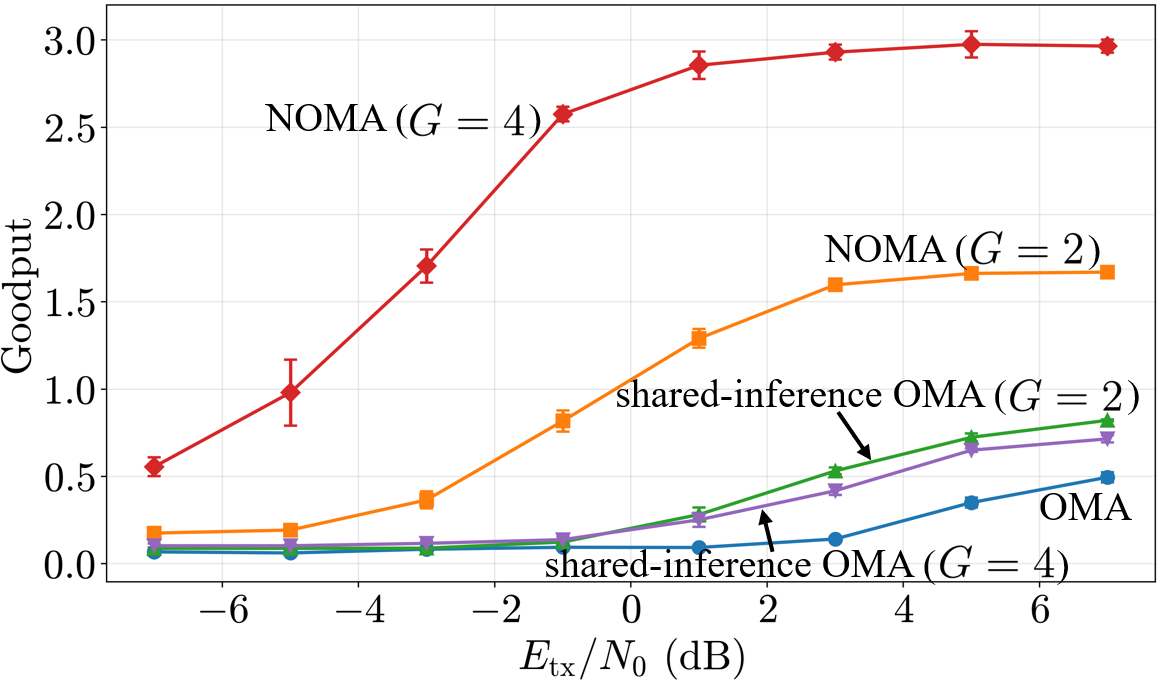}
    \caption{Goodput $\mathcal{T}$ versus $E_{\rm tx}/N_0$ for OMA-NC and NOMA-NC with group sizes $G=2$ and $G=4$ for the DVS128 Gesture dataset.} \label{awgn_snrg}
\end{figure}

For each test sample we compute the realized normalized transmission energy
$E_{\mathrm{tx}}/N_0$, i.e., the per-device per-subcarrier transmission energy relative to the noise power, together with the receiver-side compute energy $E_{\mathrm{rx}}$ in \eqref{rxe}, and summarize the resulting empirical densities over the test set. As shown in Fig. \ref{hist}, for the N-MNIST dataset, the distributions of the ratio $E_{\mathrm{tx}}/N_0$ shift with the group size, concentrating around 3.5 dB for the $G=4$ schemes and around 4.5 dB for the $G=2$ schemes, with the OMA baseline lying in between. For a given group size $G$, the NOMA-NC and shared-inference OMA-NC distributions coincide, since the transmission energy is governed by the spike sparsity and is therefore unaffected by whether the $G$ devices transmit on shared or on disjoint resources. The receiver computing energy $E_{\rm rx}$ likewise coincides for NOMA-NC and shared-inference OMA-NC at each group size $G$, since both decode all $G$ devices through a single shared inference pass and thus amortize the same number of synaptic operations over the group. Increasing the group size from $G=1$ (OMA) to $G=2$ and $G=4$ shifts the computing energy distribution from a mean of about $9 \mu J$ down to about $8$ and $4.5$ $\mu J$, respectively, and renders it substantially more concentrated, indicating that the per-device computational cost becomes increasingly predictable as more devices share the inference.

Fig.~\ref{awgn_snr} reports the corresponding goodput $\mathcal{T}$ as a function of the ratio $E_{\mathrm{tx}}/N_0$. NOMA-NC is seen to deliver substantially higher goodput than OMA across the entire range of values of the transmit energy, with the gap widening as the group size $G$ increases. At
$E_{\mathrm{tx}}/N_0= 7$~dB, OMA-NC saturates at $\mathcal{T}= 0.6$,
whereas NOMA-NC reaches
$\mathcal{T}= 1.27$ for $G=2$ and $\mathcal{T}= 2.45$ for $G=4$,
scaling the goodput by approximately a factor equal to the group size $G$. By contrast, the shared-inference
OMA-NC curves for $G=2$ and $G=4$  track the OMA-NC baseline,
confirming that this goodput gain arises from the over-the-air bundling rather
than from the shared inference alone. At low ratio $E_{\mathrm{tx}}/N_0$,
the gain is smaller in absolute terms but consistently present, indicating that
the over-the-air superposition retains its goodput advantage even when
classification is made more challenging by noise.

As shown in Figs.~\ref{histnmnist} and~\ref{awgn_snrg}, the same conclusions carry over to the more challenging DVS128 Gesture benchmark. In Fig.~\ref{histnmnist}, the receiver-side energy $E_{\rm rx}$ falls from a mean of roughly $120$ $\mu$J for OMA to $50$ $\mu$J for $G=2$ and $22$ $\mu$J for $G=4$, the $E_{\rm tx}/N_0$ distributions shift to lower values as group size grows, and the NOMA-NC and shared-inference OMA-NC distributions coincide at each group size, with the larger values relative to Fig.~\ref{hist} reflecting the deeper and wider DVS128 decoding SNN. The goodput in Fig. \ref{awgn_snrg} rises from 0.6 for OMA to 1.65 for $G=2$ and to 3 for $G=4$ while the shared-inference OMA-NC curves track the OMA baseline, so the goodput again scales at least linearly with the group size across the full range of $E_{\rm tx}/N_0$.

\subsection{Hypernetwork-Based Channel Adaptation}
We now remove channel knowledge from the devices and evaluate the proposed Hyper-NOMA-NC, in which a hypernetwork performs receiver-side equalization from the pilot as
described in Section~V. We compare this approach against the hypernetwork-equipped OMA baseline, as well as against the joint end-to-end training benchmark in both OMA and NOMA modes. 

\begin{figure}[t!]
    \centering

        \includegraphics[width=0.6\linewidth]{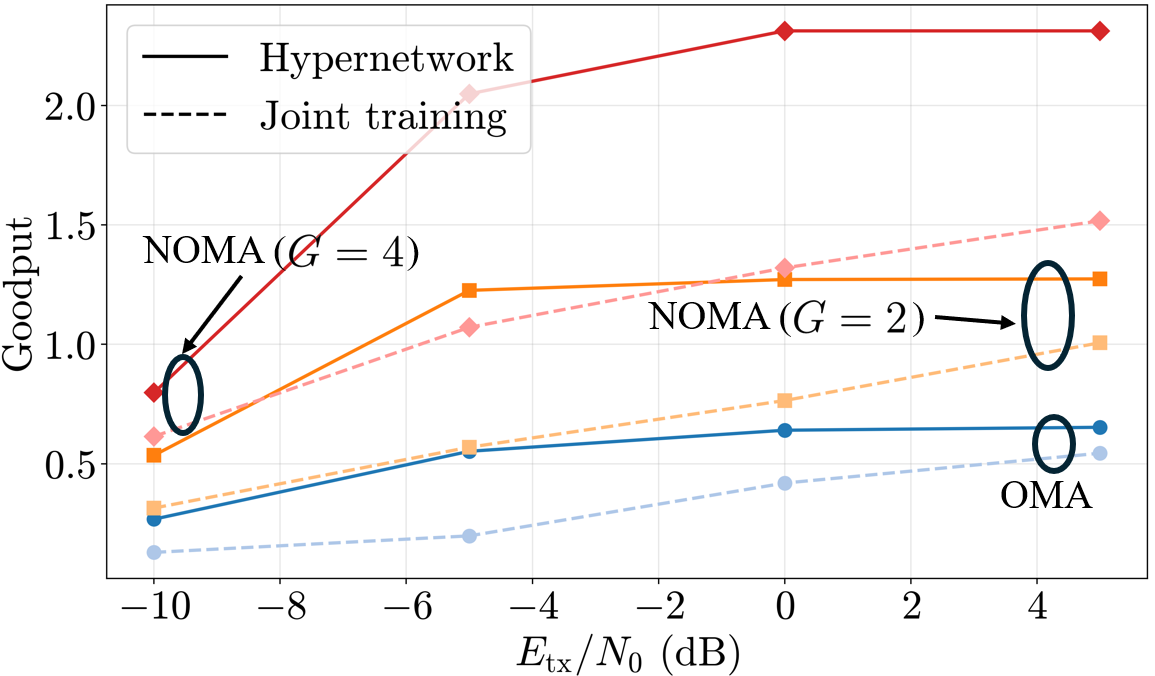}
    \caption{Goodput $\mathcal{T}$ versus $E_{\rm tx}/N_0$ for OMA-NC and  NOMA-NC with group sizes $G=2$ and $G=4$ for the N-MNIST dataset without channel knowledge at the devices.} \label{snr_fading}
\end{figure}

Fig.~\ref{snr_fading} reports the goodput as a function of $E_{\rm tx}/N_0$. For clarity, the shared-inference OMA curves are omitted as they closely track the OMA baseline in goodput and would clutter the figure. Hyper-NOMA-NC consistently outperforms joint end-to-end
training, with the gap widening as the group size $G$ grows. At high $E_{\mathrm{tx}}/N_0$,
Hyper-NOMA-NC reaches $\mathcal{T}= 1.27$ for $G=2$ and
$\mathcal{T}= 2.3$ for $G=4$, against $\mathcal{T}= 1$ and
$\mathcal{T}= 1.5$ for the corresponding joint-training variants, while the
OMA baselines remain low at $\mathcal{T}= 0.6$ with the hypernetwork and
$\mathcal{T}= 0.5$ with joint training. Hyper-NOMA-NC's advantage is
most pronounced at $G=4$, where the stronger inter-device interference makes
pilot-driven, channel-conditioned normalization most valuable.

\begin{figure}[t!]
    \centering
    \includegraphics[width=1.0\linewidth]{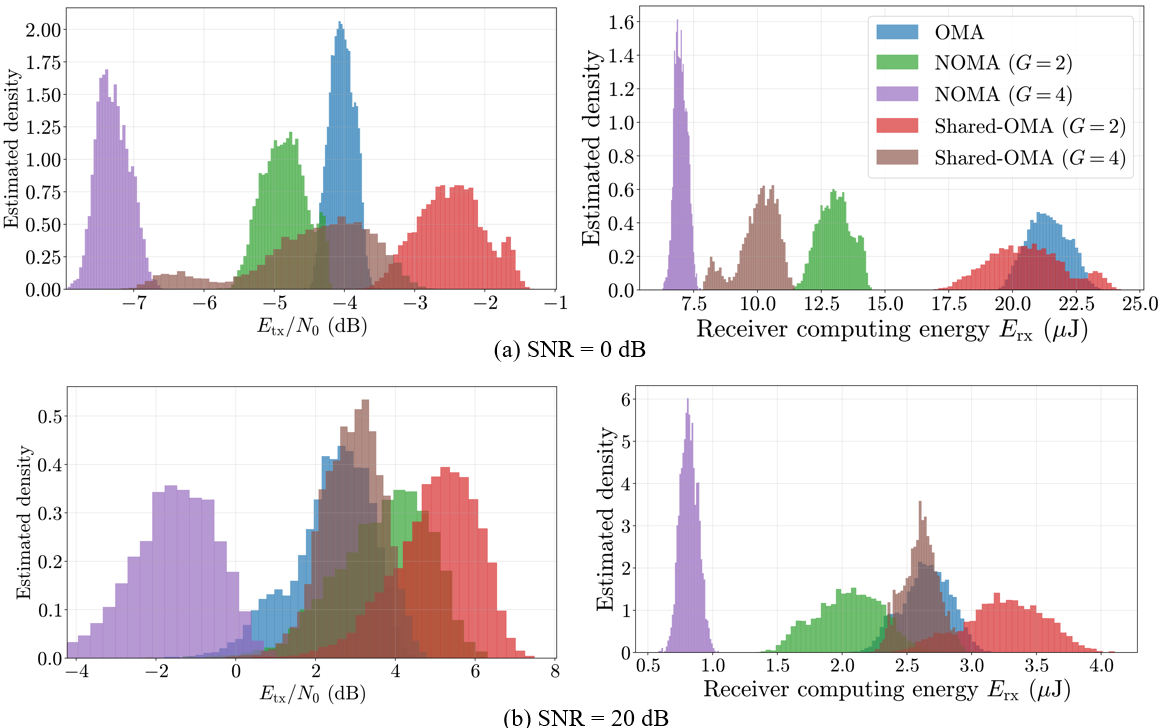}
    \vspace{-1cm}
    \caption{Left: Empirical distribution of the ratio $E_{\rm tx}/N_0$ between transmission energy and noise power. Right: Empirical distribution of the receiver computing energy $E_{\rm rx}$ in \eqref{rxe}. Results are shown for OMA-NC and NOMA-NC with $G \in \{2, 4\}$ for the N-MNIST dataset without channel knowledge at the devices.
} \label{histf}
\end{figure}

Fig.~\ref{histf} reports the transmission and receiver-side energy distributions at SNR values of 0 and 20 dB. We report only the hypernetwork-equipped schemes here, since Fig.~\ref{snr_fading} already establishes that the hypernetwork outperforms joint training in goodput, making it the preferred choice for the energy comparison. As under perfect compensation, the distribution of $E_{\rm tx}/N_0$ shifts to lower values as the group size $G$ grows, with the OMA baseline lying among the $G=2$  schemes. Unlike the perfect-compensation case, however, the NOMA-NC and shared-inference OMA-NC distributions no longer coincide for a given group size $G$. Because the two schemes equalize the channel differently, with NOMA-NC operating on a superposed pilot and shared-inference OMA-NC on clean per-device pilots, they induce different firing activity in the decoding SNN, and hence differ in both $E_{\rm tx}/N_0$ and the receiver-side energy $E_{\mathrm{rx}}$. Nonetheless, the receiver-side energy $E_{\mathrm{rx}}$ still falls substantially with the group size for both schemes, although the reduction is sub-proportional to the group size $G$, since the shared decoding pass is amortized across the group while the per-device unbinding is applied to each device individually. 

\begin{figure}[t!]
    \centering

        \includegraphics[width=0.6\linewidth]{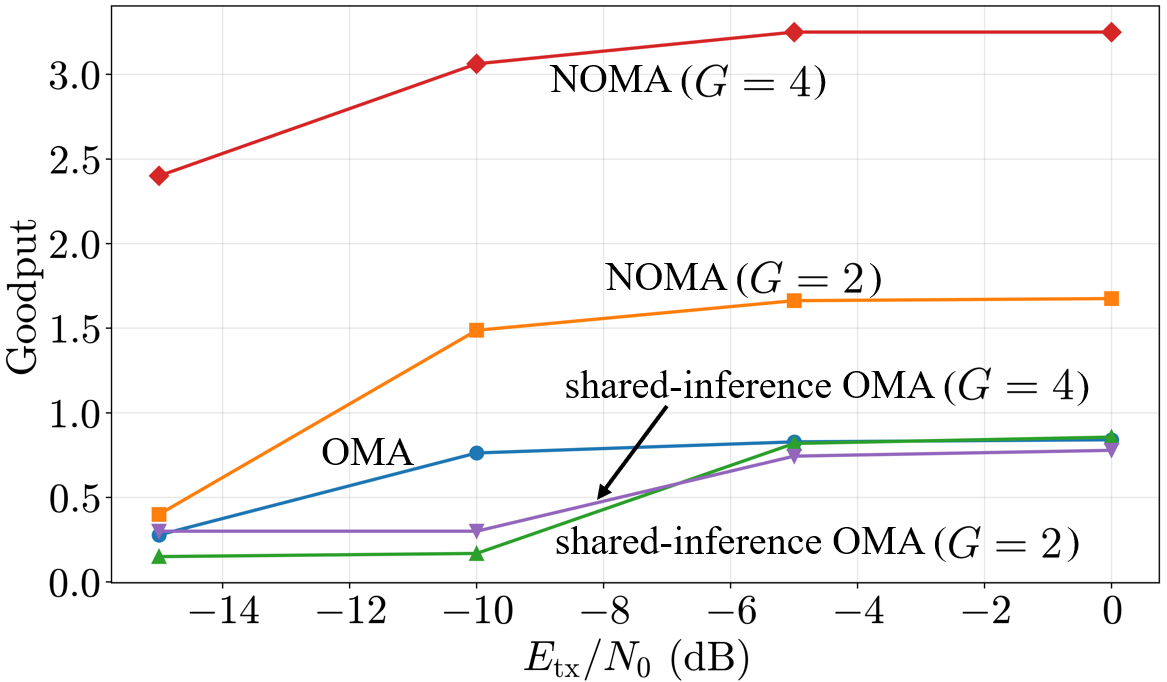}
    \caption{Goodput $\mathcal{T}$ versus $E_{\rm tx}/N_0$ for OMA-NC and  NOMA-NC using hypernetworks with group sizes $G=2$ and $G=4$ for the DVS128 Gesture dataset without channel knowledge at the devices.} \label{snrg_fading}
\end{figure}

\begin{figure}[t!]
    \centering
    \includegraphics[width=1.0\linewidth]{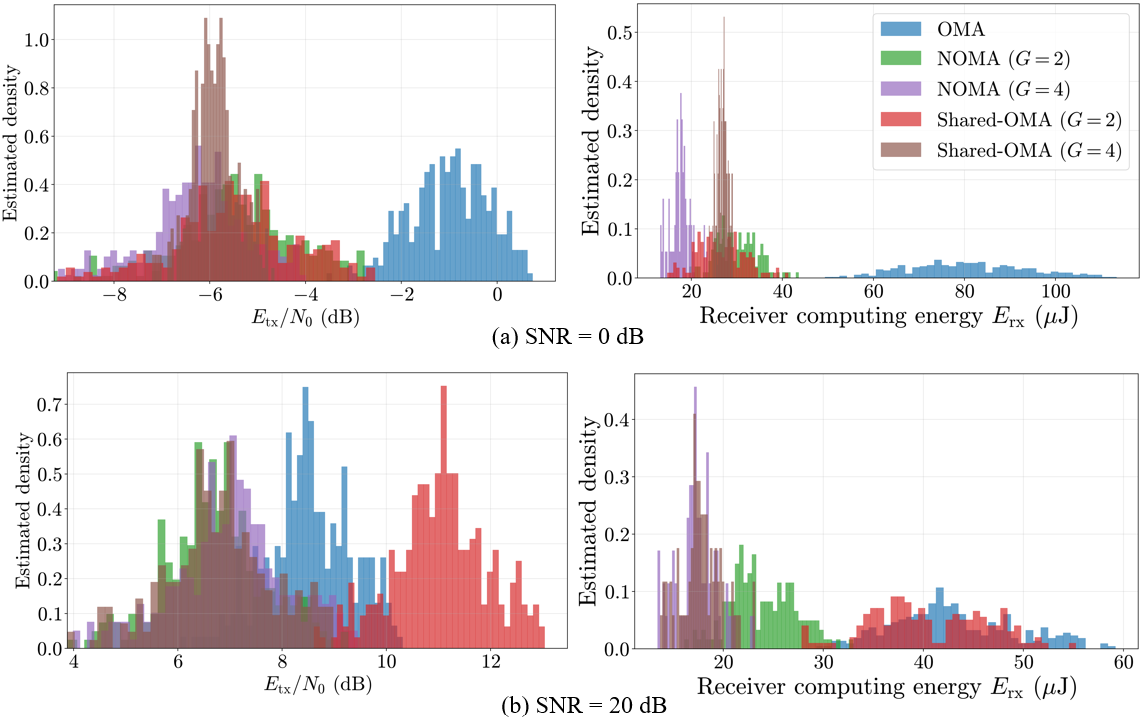}
    \vspace{-1cm}
    \caption{Left: Empirical distribution of the ratio $E_{\rm tx}/N_0$ between transmission energy and noise power. Right: Empirical distribution of the receiver computing energy $E_{\rm rx}$ in \eqref{rxe}. Results are shown for OMA-NC and NOMA-NC with $G \in \{2, 4\}$ for the DVS128 Gesture dataset without channel knowledge at the devices.
} \label{histg}
\end{figure}

As shown in Figs.~\ref{snrg_fading} and~\ref{histg}, the same trends carry over to the DVS128 Gesture benchmark without channel knowledge at the devices. Hyper-NOMA-NC reaches $\mathcal{T}=1.5$ for $G=2$ and $\mathcal{T}=3.3$ for $G=4$ at high $E_{\rm tx}/N_0$, while the shared-inference OMA curves track the OMA baseline. In Fig.~\ref{histg}, both $E_{\rm tx}/N_0$ and the receiver-side energy $E_{\rm rx}$ fall with the group size $G$ at SNR values of 0 and 20 dB.

\section{Conclusions}
This paper proposed NOMA-NC, a VSA-based non-orthogonal multiple-access neuromorphic protocol for parallel remote inference, in which devices bind their sparse spike feature maps with device-specific permutation keys and transmit concurrently, so that the over-the-air superposition realizes VSA bundling and a shared decoding SNN recovers all decisions in a single inference pass. A hypernetwork variant, Hyper-NOMA-NC, removes the need for transmitter-side channel inversion. On N-MNIST and DVS128 Gesture, both schemes scale the goodput approximately linearly with the group size $G$ while sharply reducing the receiver-side computing energy and preserving per-device transmission energy. Future work includes extending NOMA-NC to heterogeneous per-device tasks and adaptive group formation that balances goodput against inter-device interference.


\small{
\bibliographystyle{ieeetr}
\bibliography{references}
}

\end{document}